# Interplay between Cu diffusion and bonding anisotropy on the thermoelectric performance of double cation chalcohalides $CuBiSeX_2$ (X = Cl, Br)

Manivannan Saminathan[a], Prakash Govindaraj[b], Hern Kim[b], Kowsalya Murugan[a]

Kathirvel Venugopal[a, *]

[a] Department of Physics and Nanotechnology, SRM Institute of Science and Technology, Kattankulathur, Tamil Nadu, 603 203, India.

[b] Department of Energy Science and Technology, Environmental Waste Recycle Institute, Myongji University, Yongin, Gyeonggi-do 17058, Republic of Korea.

*Corresponding Author E-mail: kathirvv@srmist.edu.in

**Abstract**

Double cation chalcohalide have recently been emerged as the interesting candidates for sustainable energy conversion applications, owing to their intrinsic chemical tunability, suitable band gap, and low thermal conductivity. With this motivation, the current study is designed to explore the structural, electron and phonon transport mechanism, and thermoelectric properties of $CuBiSeX_2$ (X = Cl, Br) through density functional theory-based computations. The experimental feasibility of the compounds is ensured, and they are predicted to be thermally, dynamically, and mechanically stable. The distinct structural attributes coupled with suitable electronic band structure promotes the electron transport properties. Comprehensively, the delocalized Cu atom enhancing the phonon scattering process and the off-centred displacement of cations leading to bonding anharmonicity results ultra-low lattice thermal conductivity ($\kappa_L$). Among these systems, $CuBiSeCl_2$ exhibits low $\kappa_L$ (0.24 W m$^{-1}$ K$^{-1}$ at 300 K) and superior thermoelectric performance ($zT$ = 1.18 at 600 K), whereas $CuBiSeBr_2$ ($\kappa_L$ = 0.65 W m$^{-1}$ K$^{-1}$ at 300 K, $zT$ = 0.68 at 600 K) demands further optimization. Overall, the study sheds light into the interplay between the Cu diffusion and bonding anisotropy in phonon propagation and establishes the potential of double-cation chalcohalides for mid-temperature thermoelectric applications.

**Graphical Abstract**

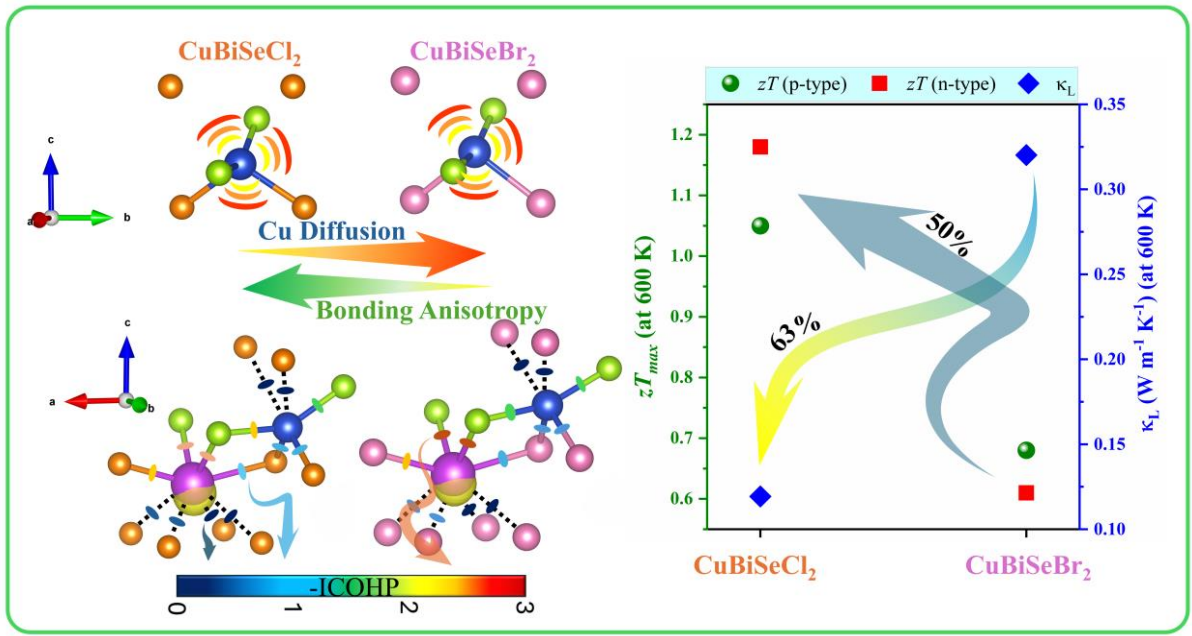

# 1. Introduction

Progress towards sustainable environment highly relies on the development and consumption of renewable energy sources.[1] In this line, utilization of waste heat can be the simultaneous solution to address the raising global warming and energy crisis. Owing to its low maintenance requirements, emission-free operation and environmentally benign nature, thermoelectricity represents an appealing energy conversion technology capable of directly converting heat into electrical energy and the other way round.[2] Generally, the metric of thermoelectric (TE) materials is labelled by the dimensionless figure-of-merit ( $zT = (\sigma S^2/\kappa_e + \kappa_L)T$ ). Here, $\sigma$, $S$, $\kappa_e$ and $\kappa_l$ denotes the material's electrical conductivity, Seebeck coefficient, thermal conductivity contributed from electrons and phonons. Mathematically, achieving higher $zT$ requires large $PF$ ($\sigma S^2$) and low thermal conductivity ( $\kappa = \kappa_e + \kappa_L$ ). Typically, materials with $zT$ more than *one* is recognized as promising thermoelectric candidate. However, ending up in high $zT$ is naturally limited by the complex relationship between $\sigma$, $S$ and $\kappa_e$.[3,4] Besides these complexities, several techniques are being established to boost their TE performance. These strategies primarily aim to optimize the $PF$ and supress the phonon propagation. The former case highly depends on the characteristics of electronic band structure, which is tuned by band engineering approaches, particularly band alignment, band convergence, band sharpening, and resonant level doping.[5–9] The latter case is handled by introducing the crystal defects, implementing nano-structuring, grain boundary engineering and so on.[10,11]

Apart from these approaches, remarkable research progress has been achieved from materials design perspective. To specifically mention, the 8-18 rule applied Heusler alloys, vacancy-filled skutterudites, chalcogenides, and clathrates are interesting.[12–14] Within this context, chalcogenides represent a fascinating family of materials due to their soft bonding nature, suitability for near-room-temperature operation, and elemental abundance. Although numerous chalcogen-containing material classes have been reported, layered chalcogenides and chalcohalides have emerged as promising candidates for thermoelectric energy conversion, owing to their optimized power factors and intrinsically low lattice thermal conductivity. These layered compounds exhibit strong intralayer bonding, while the end-to-end layers are connected through weak van der Waals forces or ionic bonds, collectively leading to favourable transport properties. Such features offer a compelling avenue to develop alternative materials that could replace conventional TE candidates like PbTe, $Bi_2Te_3$, and GeTe, which are limited by elemental scarcity and sustainability concerns[15–17].

Comprehensively, the basic search relies in discovering materials with low thermal conductivity without resisting the electron transportation. Recently, mixed anion compounds with two or more anion within the single phase is found to exhibit low $\kappa$. The inherent bonding heterogeneity accompanied by cation bonding to more than one anion facilitates in forming locally distorted structures.[18] The thermoelectric properties of this class of materials remains less explored. Some of the explored interesting systems includes, $Bi_{13}S_{18}Br_2$, $CuBiSCl_2$ and BiCuSeO. The n-type $Bi_{13}S_{18}Br_2$ and $CuBiSCl_2$ exhibits maximum $zT$ of 1.0 at 748 K ($\kappa_L$ = 0.55 W m$^{-1}$ K$^{-1}$ at ~298 K) and 1.57 at 700 K ($\kappa_L$ = 0.40 W m$^{-1}$ K$^{-1}$ at 300 K).[19,20] On the other hand, p-type BiCuSeO shows maximum $zT$ of 0.7 with lattice thermal conductivity of 0.5 W m$^{-1}$ K$^{-1}$ at 773 K.[21] It is worth noting that $CuBiSCl_2$ with mixed anion accompanied by double cation possess higher thermoelectric performance.

That is, the incorporation of double cation, specifically Cu and Bi further promotes the phonon scattering process and satisfies the "Phonon Glass Electron Crystal" paradigm. Earlier report clearly states that the diffusion of Cu atoms accompanied by structural anisotropy, bonding heterogeneity and 6s$^2$ lone pair electrons strongly enhances the thermoelectric performance of $CuBiSCl_2$.[22] Cara J. Hawkins et al., revealed the successful formation of $CuBiSeCl_2$ compound by replacing the chalcogen atom. The study on the optical and thermal transport properties of $CuBiSeCl_2$ found the band gap ($E_g$) and thermal conductivity to be 1.33 eV and 0.27(4) W m$^{-1}$ K$^{-1}$.[23] Later, Yuzhou Hao et al. exposed the occurrence of Cu delocalization in $CuBiSeCl_2$ and correlated it as the origin for the ultralow thermal conductivity.[24] Although this compound fulfils the key criteria of efficient thermoelectric materials, namely, a narrow $E_g$ with ultralow $\kappa_L$, their thermoelectric properties have not yet been systematically explored, thereby motivating the present investigation.

In this work, the density functional theory is utilized to comprehensively understand the unique structural and bonding characteristics of $CuBiSeX_2$ (X = Cl, Br) compounds. The possibility of forming the targeted compounds through experiments and their stability in various aspects is ensured. The presence of 6s$^2$ lone-pair electrons and heterogeneous bonding exerts a pronounced impact on the transport properties. The existence of suitable electronic band gap and band structure is also regarded as an added advantage. The results of molecular dynamics simulations, flatter potential energy surface, larger atomic displacement parameter coupled with large bonding anharmonicity yields low lattice thermal conductivity in both the systems. These intrinsic peculiarities cumulatively give rise to negligible $\kappa_e$ without affecting the electronic conductivity, especially in CuBiSeCl$_2$. As a consequence of the "Phonon glass electron crystal" behaviour, superior thermoelectric performance is observed in CuBiSeCl$_2$. Collectively, this study not only provides structure-property relationship, but also calls for the experimental exploration of low cost and sustainable double cation chalcohalides for thermoelectric applications.

2. **Computational Methodology**

First-principles calculations based on density functional theory (DFT) were performed using the Vienna Ab initio Simulation Package (VASP).[25,26] The interactions between core and valence electrons were considered using the projector augmented wave (PAW) method.[27] The Perdew–Burke–Ernzerhof (PBE) functional within the generalized gradient approximation (GGA) was employed to describe the exchange-correlation effects.[28] The converged plane-wave energy cutoff of 400 eV and 11 × 9 × 3 $k$-mesh within Monkhorst-Pack scheme was used. The van der Waals (vdW) interactions within the crystal was treated with the DFT-D3 correction with Becke–Johnson damping (IVDW = 12).[29] A Hubbard correction with an effective potential of $U_{eff}$ = 6 eV ($U_{eff}$ = U + J) was introduced to treat the localized Cu 3d orbitals.[30] The suitability of the chosen $U_{eff}$ value was verified through a comparative analysis, as presented in the Supporting Information (refer **Figure S1**). The structures were fully relaxed upto $10^{-8}$ eV total energy and $10^{-7}$ eV Å$^{-1}$ Hellmann–Feynman forces. Ab initio molecular dynamics (AIMD) simulations were carried out using the Nose–Hoover thermostat (MDALGO = 2) within NVT ensemble.[31,32] The structures were initially equilibrated for 1 $p$s in steps of 2 fs, continued by a production run of 6 $p$s. Mechanical stability was evaluated by calculating the elastic constants via finite deformation method in the ElaStic code.[33] The bonding characteristics were analyzed by the LOBSTER code.[34]

The electron transport properties were computed using the AMSET code by solving the linearized Boltzmann transport equation (LBTE) based on the BoltzTraP2 framework.[35,36] This approach enables an effective evaluation of electron scattering processes and their corresponding relaxation times. In this work, three primary scattering mechanisms namely acoustic deformation potential (ADP), ionized impurity (IMP), and polar optical phonon (POP) scattering. To incorporate these scattering mechanisms, several supporting calculations were performed, which includes density functional perturbation theory (DFPT) computations to derive polar phonon frequencies, dielectric constants, and elastic constants. The calculated electronic band structure using a dense $k$-mesh was subsequently employed to solve the LBTE.

Detailed implementation procedures can be found in Ref. 32. The obtained parameters were used to calculate the electron transport properties, and the convergence tests of the $S$ and $\sigma$ with respect to the interpolation factor are provided in the **Table S1** and **Figure S2**.

Furthermore, $\kappa_L$ was calculated using the modified Debye–Callaway (mDC) model in the AICON2 code.[37] To solve the mDC model, phonon dispersion curves were computed at three different volumes such as equilibrium, slightly expanded, and slightly compressed volumes. All phonon calculations were done through the Phonopy code based on the finite displacement approach.[38,39] To ensure the accuracy of the phonon dispersion spectra, the simulation cell was expanded in all crystallographic directions to achieve a dimension of at least 1 nm. Specifically, a 2 × 3 × 1 supercell with 120 atoms was employed. Owing to the orthorhombic symmetry of the primitive cell (20 atoms per unit cell), 20 displaced configurations were generated. Single-point energy calculations were executed for the displaced structure to obtain the second-order interatomic force constants (IFCs). Using the computed IFCs, phonon dispersion, phonon density of states (PDOS), group velocity, and Grüneisen parameters were determined.

### 3. Results and Discussions
#### 3.1. Structure and Bonding Analysis

The complex double cation chalcohalides $CuBiSeX_2$ (X = Cl, Br) crystallizes in orthorhombic *pnma* space group (No.: 62). The crystal structure of $CuBiSeCl_2$, and $CuBiSeBr_2$ with four-unit formula units per unit cell, generated using VESTA is shown in **Figure 1(a, b)**, respectively.[40] The relaxed lattice parameter of the studied compounds listed in **Table S2**, has good agreement with the earlier reports.[23,41] The unit cell constitutes of $CuSe_2X_2$ tetrahedra and $BiSe_2X_6$ polyhedra. The $Cu^+$ cation has *four* co-ordination environments with *two* Cu—Se bonds and *two* Cu—Cl bonds, while $Bi^{3+}$ has *eight* co-ordination with *two* Bi—Se bonds and *six* Bi—Cl bonds. These two Cu and Bi centered tetrahedra and polyhedra are stacked as layers within the cell along *z*-direction. Within this structure, the tetrahedron gets connected to two neighboring polyhedral by edge sharing of X-Se edges. The detailed structural peculiarities in comparison with similar double cation chalcohalides are described in ref. 41.

The chemical bonding character within the systems is studied using the electron localization function (ELF). **Figure 1(c, d)** represents the 2D ELF of CuBiSeCl$_2$, and CuBiSeBr$_2$ respectively. Here, blue, green, and red colors correspond to ELF = 0 (no localization), ELF = 0.5 (equal distribution of electrons) and ELF = 1 (complete localization).[42,43] Irrespective of the compounds, the ELF between Bi and Se is around 0.5, indicating the mutual sharing of electrons among the atoms, leading to the formation of covalent bond. In both these systems, Cu atoms are delocalized with ELF value closer to *zero*. The significance of the presence of Cu delocalization is discussed in Section 3.5.1. On the other hand, Bi—X bond becomes comparatively stronger, as the Cl atom is replaced by Br. The mushroom shaped (red color) ELF shows the lone pair electrons of Bi atoms (refer **Figure S3**). By swapping the Cl atom by Br, the lone pair electrons get tilted. This behavior can be attributed to the strong electronegativity of the halogen atoms, that electrostatically repels the lone pair electrons.

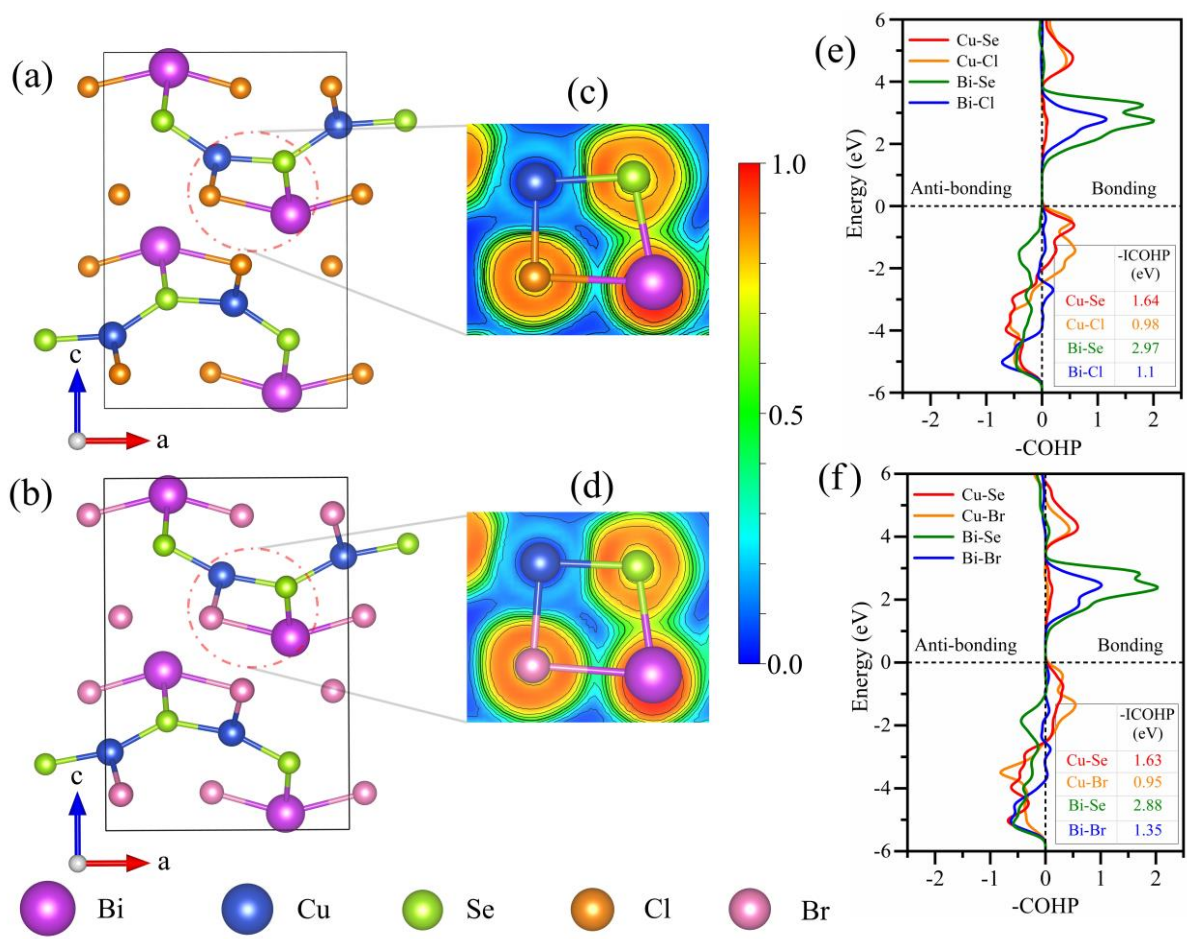

**Figure 1. (a, b)** Crystal structure, **(c, d)** 2D electron localization function, and **(e, f)** Crystal Orbital Hamilton Population of CuBiSeCl$_2$ and CuBiSeBr$_2$.

Further, the quantification of the bond strength is done by performing projected crystal orbital Hamilton population (pCOHP) and Integrated crystal orbital Hamilton population (ICOHP) analysis, shown in **Figure 1(e, f)**. The clear bonding and antibonding states of Cu—Se, Cu—(Cl/Br) and Bi—Se bonds are observed -2 eV lower the Fermi energy. The occupancy of Bi—Se antibonding states indicates the softening of lattice, which promotes phonon scattering process, Notably, the -pCOHP values of Bi—(Cl/Br) is negligible below -2 eV, revealing the absence of covalent bonding nature and thus most-likely favors ionic bonding. Moreover, the -ICOHP values also signifies the strength of the bonds, wherein more negative -ICOHP value represents the strong bonding. In both the systems, the -ICOHP of Bi—Se bond is comparatively higher than all the bonds, suggesting their stronger bonding, while Cu—(Cl/Br) being the weakest. The higher -ICOHP values of Bi—Br bond than Bi—Cl indicates the increase in bond strength by replacing the halogen atom. Besides, the significant -ICOHP of Cu—Se bond signifies the non-negligible interaction of Cu with Se atom. As a whole, both the structures exhibit heterogeneous bonding characteristics, an indicator of low thermal conductivity.

### 3.2. Stability Analysis

To ensure the experimental synthesizability, the formation and cohesive energy of both the studied systems are calculated using the following equations:

$$E_{\text{Form}} = \frac{E(\text{CuBiSeX}_2) - E(\text{Cu}_{\text{solid}}) - E(\text{Bi}_{\text{solid}}) - E(\text{Se}_{\text{solid}}) - E(\text{X}_{2(\text{solid})})}{5} \text{ eV/atom} \quad (1)$$

$$E_{\text{coh}} = \frac{(E(\text{Cu}_{\text{atom}}) + E(\text{Bi}_{\text{atom}}) + E(\text{Se}_{\text{atom}}) + 2E(\text{X}_{\text{atom}})) - E(\text{CuBiSeX}_2)}{5} \text{ eV/atom} \quad (2)$$

Here, $E(\text{CuBiSeX}_2)$ represents the ground state energy of the compounds ( in eV/f.u.), $E((\text{Cu/Bi/Se/X})_{\text{solid}})$ is the energy of corresponding elements in its ground state structure ( in eV/f.u.), and $E((\text{Cu/Bi/Se/X})_{\text{atom}})$ is the energy of the individual atoms (in eV/atom).[44,45] The calculated formation energy of CuBiSeCl$_2$, and CuBiSeBr$_2$ is -0.56, and -0.43 eV/atom, respectively. The negative values indicates that the structure is energetically favourable and can be synthesized experimentally. Further, the cohesive energy shows the amount of energy required to breakdown the compound into individual atom. The calculated cohesive energy of CuBiSeCl$_2$, and CuBiSeBr$_2$ is 3.42, and 3.20 eV/atom, respectively. The positive values ensure the structural stability of the titled compounds. As the objective of the study is to investigate the thermoelectric performance, a temperature dependent property, it is important to confirm the thermal stability of targeted systems. The AIMD simulation of CuBiSeCl$_2$, and CuBiSeBr$_2$ at 300 and 600 K up to 6 *p*s is shown in **Figure 2(a, b)**. The negligible fluctuation in energy over the entire simulation time, even at higher temperature indicates that the systems are stable up to 600 K.

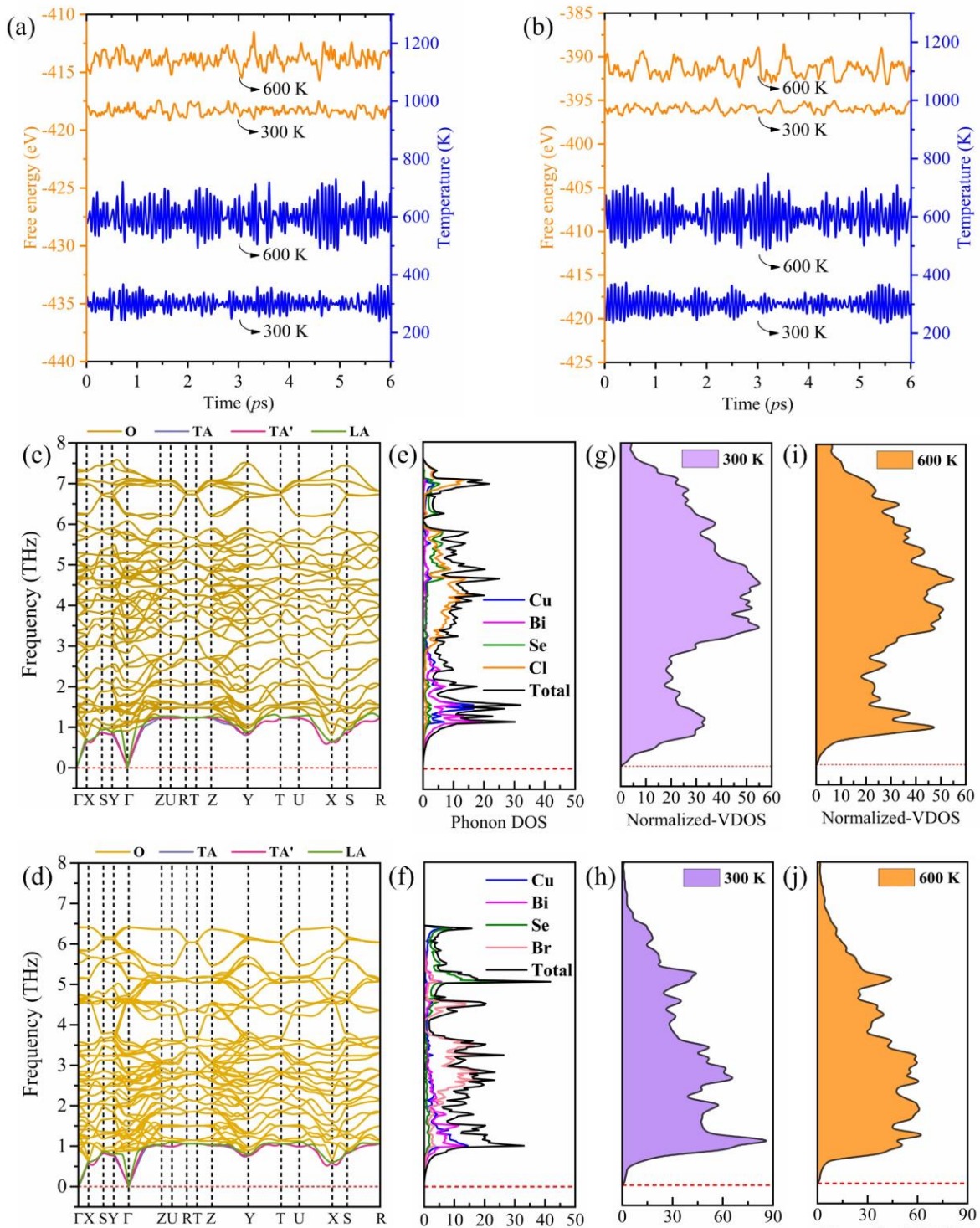

**Figure 2.** (a, b) AIMD simulation of free energy and temperature up to 6 *p*s, (c, d) phonon dispersion spectrum, (e, f) phonon density of states, Normalized-Vibrational Density of States at (g, h) 300 K and (i, j) 600 K for $CuBiSeCl_2$ and $CuBiSeBr_2$.

The dynamical stability of the titled systems is investigated from the phonon dispersion spectrum shown in **Figure 2(c, d)**. The absence of imaginary frequency over the entire BZ signifies that $CuBiSeCl_2$, and $CuBiSeBr_2$ are dynamically stable. Both the spectra have 60 phonon modes, including 3 acoustic and 57 optical branches. Generally, the heat transport in solids is primarily governed by acoustic modes. Hence, the transverse acoustic (TA, TA'), and longitudinal acoustic (LA) branches are distinguished by blue, purple and green color. The overall frequency of $CuBiSeCl_2$ (7.5 THz) is higher than $CuBiSeBr_2$ (6.5 THz). In both these systems, the acoustic and optical branches overlap around 0.5 to 1 THz. This strong coupling promotes the phonon scattering process and thus leads to reduction in heat transport. The presence of optical-optical (O-O) frequency gap is known to increase the phonon propagation in solids, which is not observed in $CuBiSeBr_2$, while $CuBiSeCl_2$ has O-O gap around 6 THz. However, this characteristic noted at higher frequency range is expected to have less effect in the overall phonon transport in $CuBiSeCl_2$. The acoustic modes along $\Gamma$-Z are larger than $\Gamma$-X, indicating the larger propagation length and thus signifies the anisotropic phonon propagation.

The PDOS shown in **Figure 2(e, f)** indicates the atoms responsible for the phonon modes. In $CuBiSeCl_2$, the acoustic modes are primarily influenced by the vibrations of Bi atoms, while the low- and high-frequency optical modes by Cu, Bi and Cl atoms. On contrary, the acoustic and low-frequency optical modes in $CuBiSeBr_2$ is contributed by Cu and Bi atoms, whereas the mid- and high-frequency optical modes by Br and Se atoms, respectively. This specifies the role of Bi and Cu atoms in determining the phonon transport phenomenon. To elucidate the temperature-dependent lattice dynamics, the vibrational density of states (VDOS) was derived from AIMD simulations performed at 300 K and 600 K for $CuBiSeCl_2$ and $CuBiSeBr_2$, as illustrated in **Figure 2(g–j)**. The nonappearance of negative-frequency components in the AIMD-derived spectra confirms the dynamic stability of both systems even at elevated temperatures. This indicates that no structural instabilities or soft-mode behaviour is raised up to 600K.

The mechanical stability of the compounds is crucial in the aspect of device manufacturing. Hence, the elastic constants are obtained from the *ElaStic* code.[46] The orthorhombic symmetry of the crystals results in *nine* independent elastic constants, listed in **Table 1**. Moreover, Born's stability criteria is used to determine the mechanical stability of the solids.[47] The necessary and sufficient condition for the orthorhombic crystal system is expressed in equation 3. Both $CuBiSeCl_2$, and $CuBiSeBr_2$ satisfies this condition and thus confirmed to be stable. Using the Voigt-Reuss-Hill approximation, the mechanical properties like Bulk modulus (*B*), shear modulus (*G*), and Young's modulus (*Y*) are calculated using equation 4.[48]

$$C_{11} > 0;\ C_{11}C_{22} > C_{12}^2;\ C_{44} > 0;\ C_{55} > 0;\ C_{66} > 0;\ C_{11}C_{22}C_{33} + 2C_{12}C_{13}C_{23} - C_{11}C_{12}^2 - C_{22}C_{13}^2 - C_{33}C_{12}^2 > 0 \quad (3)$$

$$B = \frac{B_V + B_R}{2};\ G = \frac{G_V + G_R}{2};\ Y = \frac{9BG}{3B+G} \quad (4)$$

$$\nu = \frac{3B - 2G}{2(3B + G)} \quad (5)$$

The larger bulk, and shear moduli of CuBiSeX$_2$ compound reveals their tendency to resist plastic deformation and fracture, while the higher Young's modulus ensures the rigitidy (refer **Table 1**). Pugh's ratio (*B/G*) is a critical parameter in determing the ductility of the material. In general, ductile material has *B/G* value higher than 1.75 and the material with *B/G* less than 1.75 is regarded as brittle.[49] The *B/G* of CuBiSeCl$_2$ (2.06) and CuBiSeBr$_2$ (2.13) confirm their ductile nature. In addition, Poisson's ratio ($\nu$) obtained through equation 5 can be utilized to determine the bonding nature of the compounds. The $\nu$ value below 0.15, 0.2-0.3, and 0.3-0.5 represents the covalent, ionic-covalent, and ionic bonding, respectively. In the present case, $\nu$ value is 0.29 for both the systems, indicating the mixed bonding nature of the compounds, as evident from the ELF calculations.

**Table 1.** Elastic constants and mechanical properties of CuBiSeX$_2$ (X = Cl, Br)

|  | CuBiSeCl$_2$ | CuBiSeBr$_2$ |
|---|---|---|
| *C$_{11}$* | 144.48 | 131.79 |
| *C$_{12}$* | 34.58 | 35.12 |
| *C$_{13}$* | 35.97 | 35.43 |
| *C$_{22}$* | 89.2 | 80.04 |
| *C$_{23}$* | 52.27 | 47.7 |
| *C$_{33}$* | 76.74 | 71.86 |
| *C$_{44}$* | 54.52 | 49.94 |
| *C$_{55}$* | 26.21 | 22.52 |
| *C$_{66}$* | 21.54 | 20.51 |
| **B** | 60.84 | 56.86 |
| **G** | 29.58 | 26.73 |
| **Y** | 76.37 | 69.32 |

### 3.3. Electronic Structure

The electronic band structure of CuBiSeCl$_2$ and CuBiSeBr$_2$ is shown in **Figure 3(a, b)**. The high symmetry *k*-path for orthorhombic crystal within the first Brillouin zone (BZ), ***Γ***-X-S-Y-***Γ***-Z-U-R-T-Z-Y-T-U-X-S-R is used. Both the structures are direct band gap semiconductors with valence band maximum (VBM) and conduction band minimum (CBM) centred around the ***Γ*** point. The calculated band gaps are 1.35 eV for CuBiSeCl$_2$ and 0.85 eV for CuBiSeBr$_2$, matching well with the previous reports (Refer **Table S2**). The density of states and projected density of states for the studied compounds is displayed in **Figure 3(c, d)** and **Figure 3(e, f)**. Irrespective of the compounds, the VBM is primarily contributed by Cu-*d* orbitals, whereas the CBM by Bi-*p* orbitals. This suggests the significant role of the double cation in determining the electron transport properties.

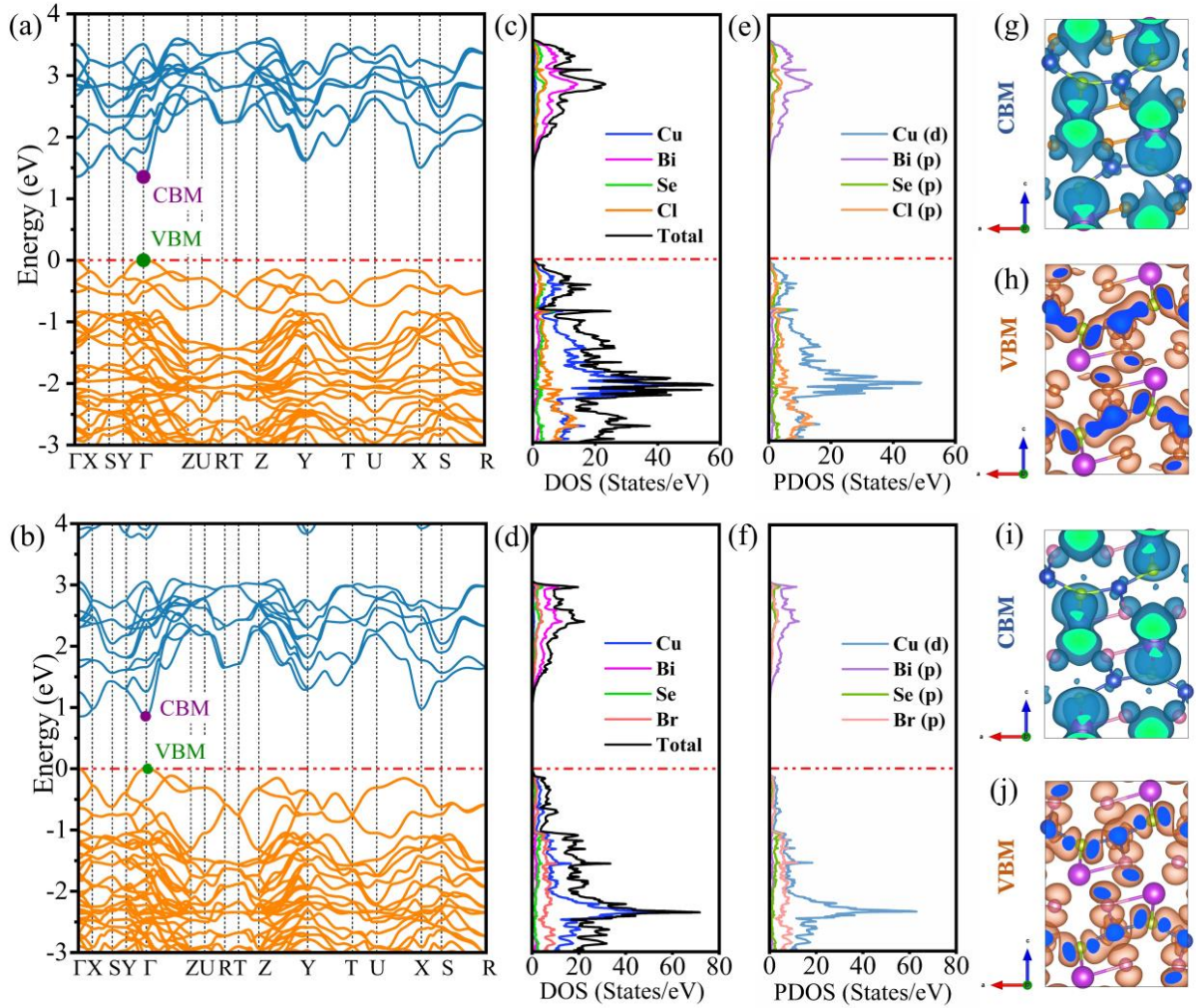

**Figure 3. (a, b)** Electronic structure, **(c, d)** electronic density of states, **(e, f)** Projected density of states for CuBiSeCl$_2$ and CuBiSeBr$_2$. Band decomposed charge densities along (010) plane around CBM and VBM of **(g, h)** CuBiSeCl$_2$ and **(i, j)** CuBiSeBr$_2$.

In semiconductors, effective mass ($m^*$) of the carriers plays a crucial role in determining the mobility and electrical conductivity. Hence, the $m^*$ is calculated using AMSET program.[50] Accordingly, the effective mass of holes ($m_h^*$) is found to be higher than electrons ($m_e^*$). This condition holds true for both the systems along $\Gamma$-Z direction (i.e.): For CuBiSeCl$_2$, $m_h^*$ (2.09 $m_0$) > $m_e^*$ (0.60 $m_0$), and for CuBiSeBr$_2$ $m_h^*$ and $m_e^*$ is 1.68 $m_0$ and 0.44 $m_0$. Here, $m_0$ is the rest mass of the electron. Mostly, the larger $m^*$ of flatter bands reduce the mobility and thus $\sigma$ of the carriers. Hence, the electrical conductivity of hole-doped system is expected to be lower than electron-doped system. Further, the band decomposed charge densities around the CBM and VBM for CuBiSeCl$_2$ and CuBiSeBr$_2$ is displayed in **Figure 3(g-j)**, respectively. For both the studied systems, the computed charge density is higher around CBM, this indicates the availability of large conduction channels for electrons to propagate, rather than hole. This also facilitates higher charge transport and thus higher electrical conductivity is expected in n-type systems.

### 3.4. Electron Transport Properties
#### 3.4.1. Relaxation Time and Electrical Conductivity

Generally, constant relaxation time approximation (CRTA) is stated to overestimate the electron transport properties. Hence, the relaxation time ($\tau$) of the carriers is calculated using advanced and reliable AMSET code in combination with VASP. This code considers various scattering mechanisms within the solids, which includes ADP, POP and IMP. The scattering rates are obtained from the deformation potential, dielectric, and elastic constants derived from VASP. Later, the Matthiessen's rule is utilized to calculate the total relaxation time of the carriers. The contribution of various scattering mechanisms to $\tau$ as a dependent of *n* and *T* or hole and electron doping in CuBiSeCl$_2$ and CuBiSeBr$_2$ is shown in **Figure. S4(a-d)**, respectively. The magnitude of $\tau$ contributed by ADP and POP remains constant with increasing *n*, whereas $\tau_{IMP}$ varies almost linearly. Under all the studied constrains, $\tau$ decreases with *T*, indicating the role of carrier scattering at higher temperature. Interestingly, POP scattering mechanism is found to be the favourable the scattering process significantly in both the systems. **Figure 4(a, b)** displays the total relaxation time against *n* and *T* for p-type and n-type doping, respectively. Irrespective of systems and type of carrier doping, $\tau$ decreases with *T*. For instance, at fixed carrier concentration (*n*) of 1 x 10$^{19}$ cm$^{-3}$, in CuBiSeCl$_2$, $\tau$ decreases from 3.86 *f*s (7.24 *f*s) at 300 K to 1.89 *f*s (2.96 *f*s) at 600 K, for p-type (n-type) doping. Whereas, in CuBiSeBr$_2$ the values changes from 4.43 *f*s (6.98 *f*s) to 2.31 *f*s (2.75 *f*s), under similar conditions. This decline can be correlated to the pronounced carrier scattering at higher temperature. Refer **Table S3** for better comparison.

The calculated electrical conductivity as a function of *n* at different temperature is displayed in **Figure 4(c, d)** and the values are listed in the **Table S4**. Interestingly, $\sigma$ increases with *n*, obeying the relation, $\sigma = ne\mu$, with $\mu$ representing the mobility of the carriers.[51] Also, $\sigma$ tends to reduce with *T*. For example, at $n = 1 \times 10^{19}$ cm$^{-3}$, for CuBiSeCl$_2$, $\sigma$ decreases from

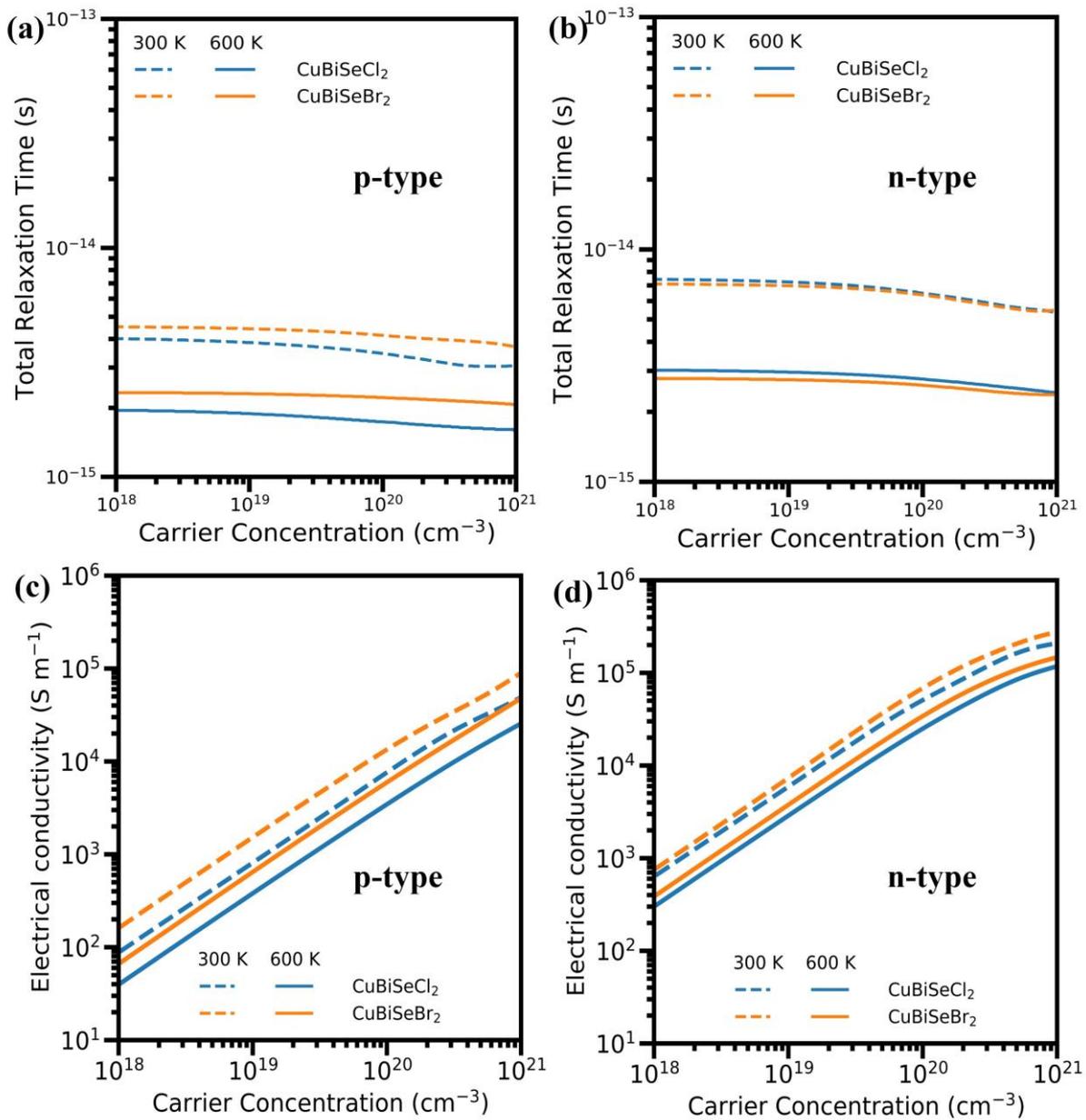

**Figure 4. (a, b)** Total relaxation time and **(c, d)** electrical conductivity with respect to carrier concentration for p- and n-type CuBiSeCl$_2$ and CuBiSeBr$_2$.

811.35 (5915.83) S m$^{-1}$ at 300 K to 380.71 (2881.56) S m$^{-1}$ at 600 K for p-type (n-type) doping. Similarly, for CuBiSeBr$_2$, in same constrain, $\sigma$ varies from 1527.39 (7336.75) S m$^{-1}$ to 645.97 (3773.11) S m$^{-1}$. This decreasing trend can be understood from the behaviour of carrier relaxation time. In comparison, n-type CuBiSeBr$_2$, exhibits higher electrical conductivity, attributed to the smaller $m^*$ of electrons accompanied by the larger conducting channels for electrons to propagate.

### 3.4.2. Seebeck Coefficient and Power Factor

The carrier concentration dependent $S$ at various temperatures for both hole and electron doped CuBiSeX$_2$ is shown in **Figure 5(a, b)** and highlights are given in the **Table S5**.

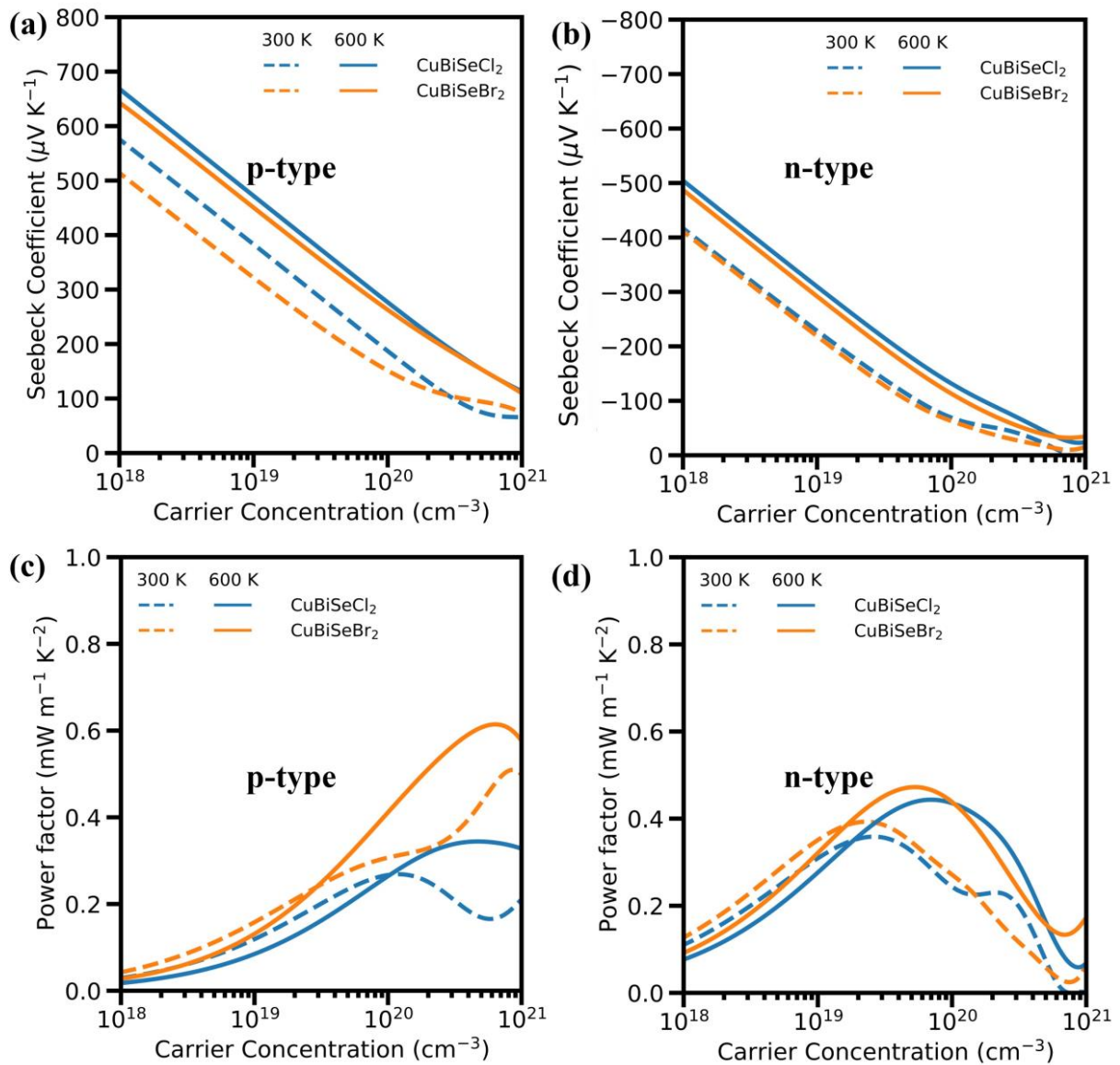

**Figure 5. (a, b)** Seebeck coefficient and **(c, d)** power factor with respect to carrier concentration for p- and n-type CuBiSeCl$_2$ and CuBiSeBr$_2$.

Obeying Mott's relation, $S = \frac{8\pi^2 k_B^2}{3eh^2} m_{DOS}^* T \left\{\frac{\pi}{3n}\right\}^{2/3}$, $S$ decreases with $n$ and increases with $T$, for both the systems, even up to heavily doped regime. Here, $k_B$ is the Boltzmann constant, $h$ is the Planck's constant, $e$ is the elementary charge, and $m_{DOS}^*$ is the density of states effective mass.[51] At $n = 1 \times 10^{19}$ cm$^{-3}$ and $T = 300$ K, $S$ for p-type (n-type) CuBiSeCl$_2$ is 383 µV K$^{-1}$ (-228 µV K$^{-1}$), whereas for CuBiSeBr$_2$, it is 321 µV K$^{-1}$ (-218 µV K$^{-1}$). With increase in temperature to 600 K, the corresponding values are 471 µV K$^{-1}$ (-309 µV K$^{-1}$) and 450 µV K$^{-1}$ (-291 µV K$^{-1}$). The observed larger $S$ for p-type systems can be assigned to the contribution from heavy holes, following inverse trend with $\sigma$ and the relation $m_{DOS}^* = N_V^{2/3} m^*$. The favorable band features led to larger $S$ in CuBiSeCl$_2$ than CuBiSeBr$_2$.

The power factor is a crucial metric to quantify the power generation ability of thermoelectric materials, obtained from the coupling effect of the $S$ and $\sigma$. The computed power factors of CuBiSeCl$_2$, and CuBiSeBr$_2$ as a function of $n$ and $T$, for both electron and hole doping are displayed in **Figure 5(c, d)**. The calculated maximum $PF$ values for CuBiSeCl$_2$, for p-type and n-type at 300 K (600 K) is 0.26 (0.34) mW m$^{-1}$ K$^{-2}$, and 0.35 (0.44) mW m$^{-1}$ K$^{-2}$, while for CuBiSeBr$_2$ it is 0.50 (0.61) mW m$^{-1}$ K$^{-2}$, and 0.39 (0.47) mW m$^{-1}$ K$^{-2}$, respectively (refer **Table S6**). The n-type CuBiSeCl$_2$ and p-type CuBiSeBr$_2$ exhibits higher $PF$ than their counterparts. This contradicting behaviour can be attributed to the inherent trade-off between $S$ and $\sigma$. For both systems and doping, at 600 K, a cross-over is observed between $10^{19}$-$10^{20}$ cm$^{-3}$, indicating the occurrence of bipolar conduction effect. This phenomenon is widely noted in narrow band gap semiconductors at higher temperatures, where both electrons and holes accounts for conduction.

### 3.4.3. Electronic Thermal Conductivity

The $\kappa_e$, quantifying the amount of heat transferred by electrons, as a function of $n$ at different temperature is given in **Figure S5(a, b)** and **Table S7**. The $\kappa_e$ has a positive correlation with $n$, following the same trend of $\sigma$, obeying Widemann-Franz law, $\kappa_e = L\sigma T$. Here, $L$ is the Lorenz number.[51] As $n$ increases, the number of carriers contributing to thermal conductivity also increases. At $n = 1 \times 10^{19}$ cm$^{-3}$, the $\kappa_e$ for hole and electron doped CuBiSeCl$_2$ at 300 K (600 K) is 0.004 W m$^{-1}$ K$^{-1}$ (0.003 W m$^{-1}$ K$^{-1}$) and 0.033 W m$^{-1}$ K$^{-1}$ (0.031 W m$^{-1}$ K$^{-1}$), whereas for CuBiSeBr$_2$ it is 0.009 W m$^{-1}$ K$^{-1}$ (0.007 W m$^{-1}$ K$^{-1}$), and 0.041 W m$^{-1}$ K$^{-1}$ (0.039 W m$^{-1}$ K$^{-1}$). A slight decrease in $\kappa_e$ is noted with increase in temperature, which can be ascribed to the carrier scattering at high temperature. Despite the observed significant values of electrical conductivity, electronic thermal conductivity remains negligible. This highlights the role of structural peculiarities in the studied double cation chalcohalides.

### 3.5. Phonon Transport Properties

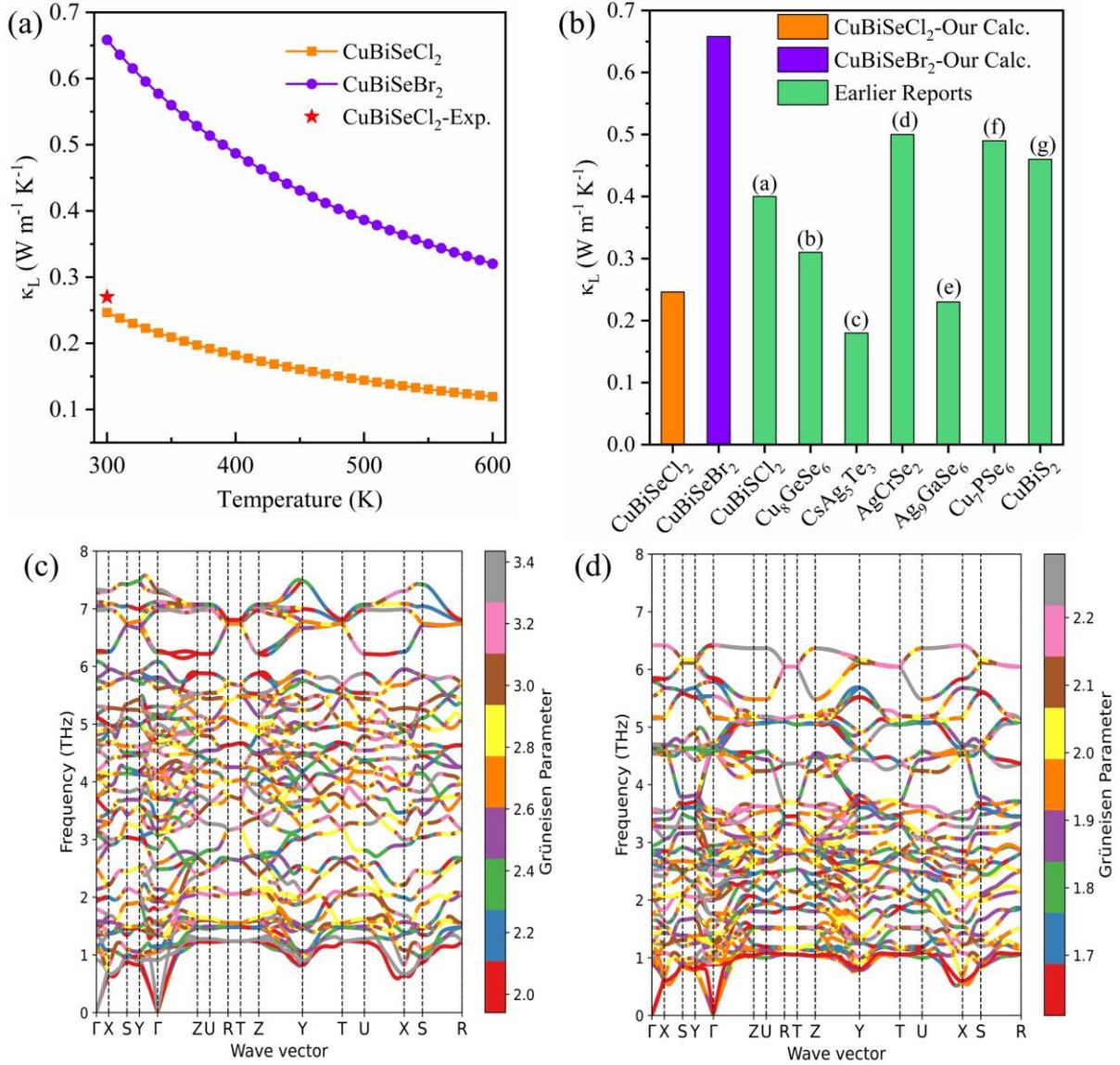

**Figure 6. (a)** Calculated lattice thermal conductivity, **(b)** Comparison of $\kappa_L$ with existing reports (a-g)[52–58] and **(c, d)** Grüneisen parameter of $CuBiSeCl_2$ and $CuBiSeBr_2$.

The temperature dependent $\kappa_L$ of both the studied systems is shown in **Figure 6(a)**. The observed decrease in $\kappa_L$ with $T$ indicates the prominent role of phonon scattering process at higher $T$. Increasing the phonon-phonon scattering decreases the mean free path of phonons, a common feature in phonon-mediated heat transfer. The calculated $\kappa_L$ for $CuBiSeCl_2$ and $CuBiSeBr_2$ at 300 K is 0.24 and 0.65 W m$^{-1}$ K$^{-1}$. This value for $CuBiSeCl_2$ agrees well with the earlier experimental result ($\kappa_L$ = 0.27 W m$^{-1}$ K$^{-1}$ at 300 K). **Figure 6(b)** compares the computed $\kappa_L$ with Cu and Ag based systems with low thermal conductivity. Interestingly, $\kappa_L$ of CuBiSeCl$_2$ is lower than CuBiSeBr$_2$, CuBiSCl$_2$, Cu$_8$GeSe$_6$, AgCrSe$_2$, Cu$_7$PSe$_6$, CuBiS$_2$ and almost equal to CsAg$_5$Te$_3$ and Ag$_9$GaSe$_6$. To understand the reason for the observed ultra-low $\kappa_L$ in CuBiSeCl$_2$, the group velocity ($v_g$) and Grüneisen parameter ($\gamma$) is computed. The $v_g$ and $\gamma$ projected phonon dispersion spectra is shown in **Figure S6(a, b)** and **Figure 6(c, d)**. The larger group velocity of CuBiSeBr$_2$ than CuBiSeCl$_2$ supports the observed larger $\kappa_L$ in Br-based system, ascribed to their direct relationship. In CuBiSeCl$_2$, the $\gamma$ of one of the acoustic branches is ~2 over the entire BZ, and the other acoustic branches have the maximum $\gamma$ of 3.4 throughout the regime, while the remaining one acoustic branch and optical branches have values ranging between 2 to 3.4. On the other hand, in CuBiSeBr$_2$, the $\gamma$ of one of the acoustic branches is ~1.6, and the other acoustic branches have a maximum of ~2, whereas the optical branches have $\gamma$ from 1.6 to 2.3. From these annotations, it is apparent that the lattice anharmonicity in CuBiSeCl$_2$ is higher than CuBiSeBr$_2$, resulting in its ultra-low $\kappa_L$. The most plausible reasons for the observed larger lattice anharmonicity can be understood by investigating the role of Cu diffusion and bonding anisotropy in determining the heat transport phenomenon of CuBiSeX$_2$ double cation chalcohalides.

### 3.5.1. Role of Cu diffusion

The atomic scale mobility of Cu ions and its influence on heat transport can be studied by analysing the AIMD results. The trajectory of tetrahedral coordination in CuBiSeCl$_2$ and CuBiSeBr$_2$ at 300 and 600 K is presented in the inset of **Figure 7(a, b)**, respectively. Although all the atoms in the tetrahedra exhibit considerable thermal vibrations, both the structures remain stable (refer **Figure S7**). Interestingly, the vibration amplitude of Cu atoms is more pronounced, indicating the possibility for the frequent movement of the respective atoms from the equilibrium position. This diffusion of Cu atoms can lead to local structural distortions and thus enhance the phonon scattering process.

The potential energy as a function of atomic displacement of atoms in CuBiSeCl$_2$ and CuBiSeBr$_2$ is shown in **Figure 7(a, b)**. This signifies the potential energy required to shift the atoms from their equilibrium position. All the atoms are observed to be in deep potential well. However, in both the systems, the potential energy surface of Cu atoms is flat, relatively flatter in CuBiSeBr$_2$, indicating the requirement of lesser energy to move the atoms from equilibrium. That is, the restoring forces are weak and thus Cu atoms are weakly bonded to the lattice. To quantitatively evaluate and compare the bond strength, the temperature dependent atomic displacement parameter (ADP) is calculated and projected in **Figure 7(c, d)**. The larger magnitude of ADP reflects the frequent vibration of the corresponding atom from its equilibrium, and thus weak bonding. In both the structures, the ADP of Bi, Se, Cl/Br atoms are less than 0.03 Å$^2$, suggesting the relatively stronger bonding of the atoms in the structure. The highest ADP of Cu atoms also support their loosely bound nature. Remarkably, the thermally induced ADP of Cu in CuBiSeBr$_2$ is larger than CuBiSeCl$_2$. This collective behavior of Cu delocalization, flatter potential energy surface, and larger ADP results comparatively higher Cu diffusion in CuBiSeBr$_2$ lattice. This characteristic affects the local structural properties and enhances the Umklapp phonon scattering process, which diminishes the phonon lifetime and mean free path. Hence, $\kappa_L$ decreases abruptly with temperature in CuBiSeBr$_2$ than CuBiSeCl$_2$.

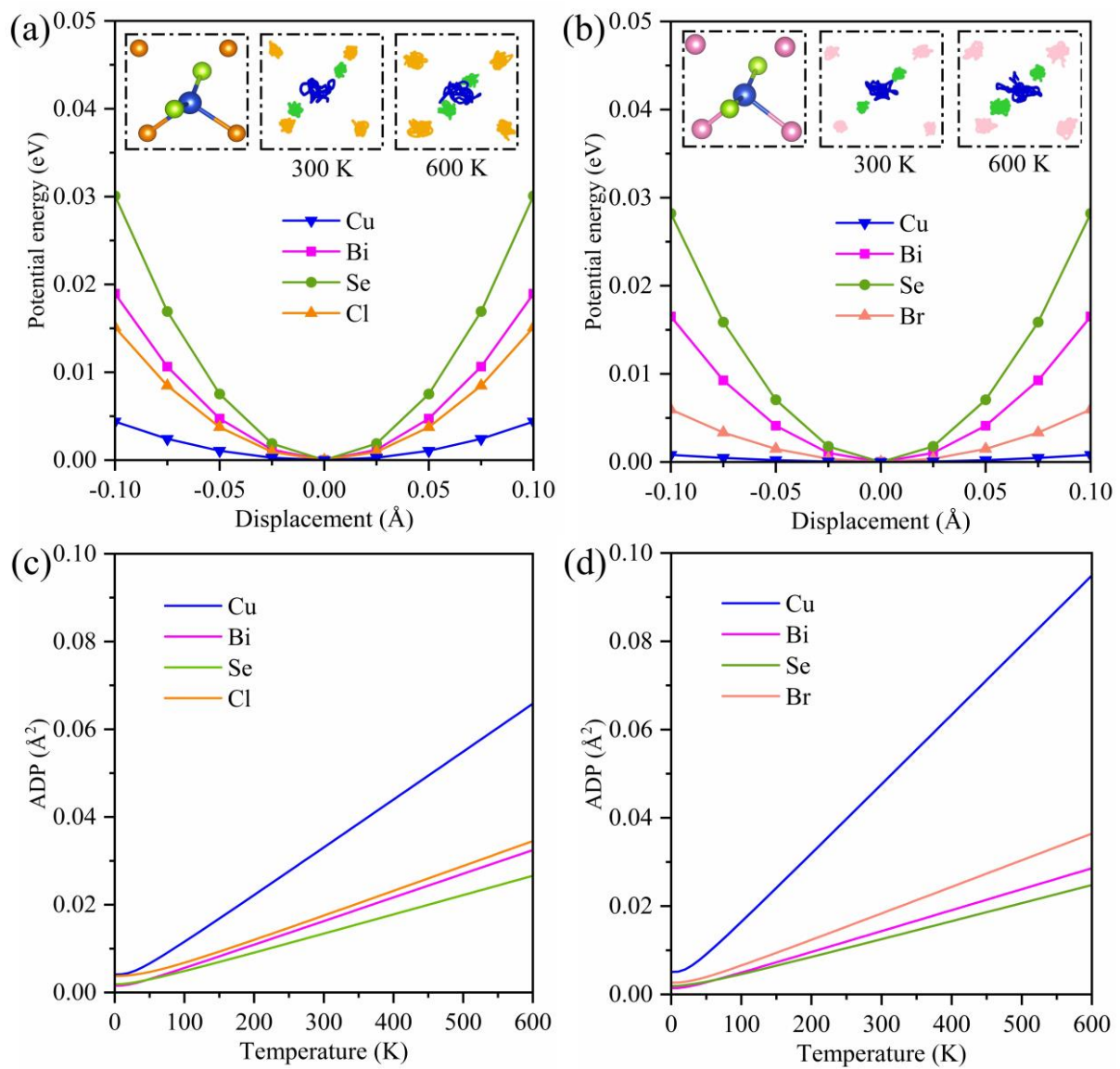

**Figure 7. (a, b)** potential energy curve and **(c, d)** atomic displacement parameter of CuBiSeCl$_2$ and CuBiSeBr$_2$. The trajectory of atoms in the CuSe$_2$Cl$_4$ and CuSe$_2$Br$_4$ octahedra is given in insets of (**a**) and (**b**) respectively.

### 3.5.2. Role of bonding anisotropy

The crystal structure discussion shows that both the systems have Cu and Bi centered tetrahedra and polyhedral environments. Here, the tetrahedra is treated as CuSe$_2$(Cl/Br)$_4$ octahedra, for better comparison with its high symmetry post-perovskite structure with *Cmcm* space group reported earlier in CuBiSCl$_2$. The Cu centered octahedra in both the structures are shown in **Figure 8(a, b)**.

It is clearly seen in both the structures that Cu atom is displaced from their central positions (i.e.): along the lower (Cl/Br)—(Cl/Br) side. To highlight the variation in bond length of octahedral environment of different systems, the percentage and magnitude of bond length is projected in **Figure 8(c, d)**. Interestingly, anisotropy is observed among the opposite Cu—Se and Cu—Cl bonds. In CuBiSeCl$_2$, the displacement of Cu atom along Cl3 and Cl4 atoms, results in reduction of Cu—Cl3 and Cu—Cl4 bond lengths than Cu—Cl1 and Cu—Cl2 correspondingly. The bond lengths of Cu—Se1 and Cu—Se2 also remains slightly anisotropic. A similar anisotropic bonding characteristic is also observed in CuSe$_2$Br$_4$ octahedra of CuBiSeBr$_2$. Comparatively, the difference in bond length between the Cu—Cl1 and Cu—Cl3 bonds is 0.86 Å (5.28 %), while that for Cu—Br1 and Cu—Br3 is 0.49 Å (3.03 %). Further, the -ICOHP values of the corresponding bonds is projected in **Figure 8(e, f)**. The distinct anisotropic features are also observed in bond strength.

Similarly, BiSe$_2$(Cl/Br)$_6$ polyhedral in CuBiSeCl$_2$ and CuBiSeBr$_2$ is shown in **Figure 8(g, h)**, respectively. The Bi atom gets displaced towards the Cl1/Br1 atom attached to the octahedra. The percentage and magnitude of bond length within the polyhedra is projected in **Figure 8(i-j)**. In both systems, the bond anisotropy is also apparent in BiSe$_2$(Cl/Br)$_6$ polyhedral, except for Bi—Se bonds. In CuBiSe(Cl/Br)$_2$, Bi—(Cl1/Br1), Bi—(Cl6/Br6) and Bi—(Cl5/Br5) bonds are shorter than Bi—(Cl2/Br2), Bi—(Cl3/Br3) and Bi—(Cl4/Br4), respectively. On the other hand, the difference in bond length between the Bi-Cl1 and Bi-Cl2 (0.31 Å, 1.29 %) is larger than Bi-Br1 and Bi-Br2 (0.19 Å, 0.76 %). The -ICOHP values in **Figure 8(k, l)** also confirms the anisotropic bond features within the polyhedral.

Overall, the bonding anisotropy in CuBiSeCl$_2$ is higher than CuBiSeBr$_2$. In CuBiSeCl$_2$, these octahedra and polyhedra is connected by Cl—Se edges, while in CuBiSeBr$_2$ it is by Br—Se edges. Interestingly, the atomic radii of Br (1.96 Å) and Se (1.98 Å) is almost similar, whereas Cl (1.81 Å) is lesser. This higher anion atomic radii difference and bonding anisotropy in CuBiSeCl$_2$ facilitates larger anharmonic vibrations within the lattice and thus larger Grüneisen parameter is observed. This in turn results lower $\kappa_L$ in CuBiSeCl$_2$ than CuBiSeBr$_2$.

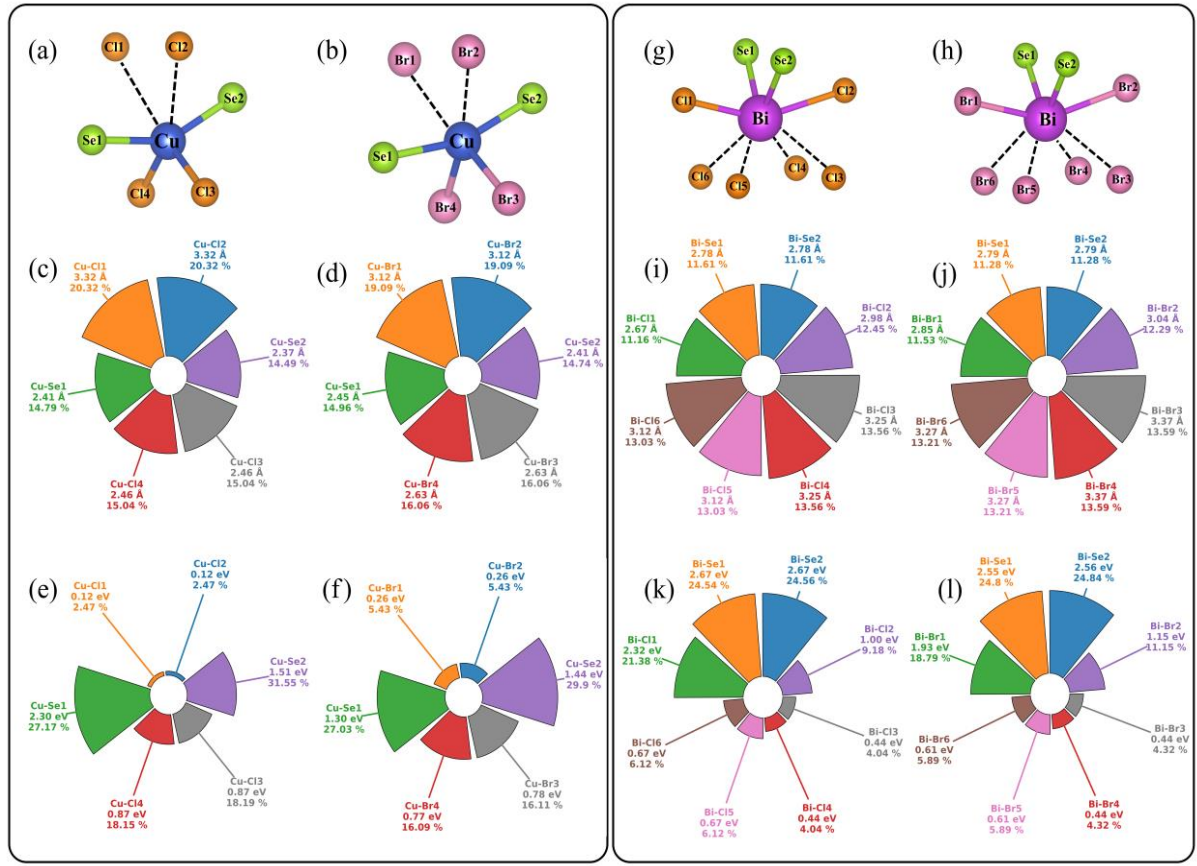

**Figure 8. (a, b)** Octahedral representation, **(c, d)** bond length and **(e, f)** -ICOHP of CuSe$_2$(Cl/Br)$_4$. **(g, h)** polyhedral representation, **(i, j)** bond length and **(k, l)** -ICOHP of BiSe$_2$(Cl/Br)$_6$. The dashed line shown in the octahedra and polyhedra is physically insignificant.

### 3.6. Figure of Merit

The calculated carrier concentration dependent $zT$ of hole and electron doped $CuBiSeX_2$ systems is displayed in **Figure 9(a, b)** and the optimum $zT$ is listed in **Table S8**. The value of $zT$ increases initially with $n$ and then decreases. The maximum $zT$ for p-type and n-type doped $CuBiSeCl_2$ at 300 K (600 K) is 0.28 (1.06) and 0.34 (1.19), while for $CuBiSeBr_2$ it is 0.13 (0.68) and 0.15 (0.62). In all the studied conditions, $zT$ increases with temperature. Interestingly, n-type $CuBiSeCl_2$ and p-type $CuBiSeBr_2$ exhibits maximum performance compared to their corresponding counterparts, as observed in *PF*. This signifies the tunability of the studied systems towards desired requirement. The calculated $zT$ is comparable with most of the Cu and Ag based thermoelectric materials like $Cu_2Se$ ($zT$ = 1.19 at 723 K), $CuSbSe_2$ ($zT$ = 0.97 at 543 K), $CuBiSCl_2$ ($zT$ = 1.57 at 700 K), $Cu_3SbSe_4$ ($zT$ = 0.80 at 650 K). $Cu_8GeSe_6$ ($zT$ = 0.50 at 750 K), $AgCrSe_2$ ($zT$ = 0.81 at 773 K), and $AgSbTe_2$ ($zT$ = 1.59 at 673 K).[20,53,59–]

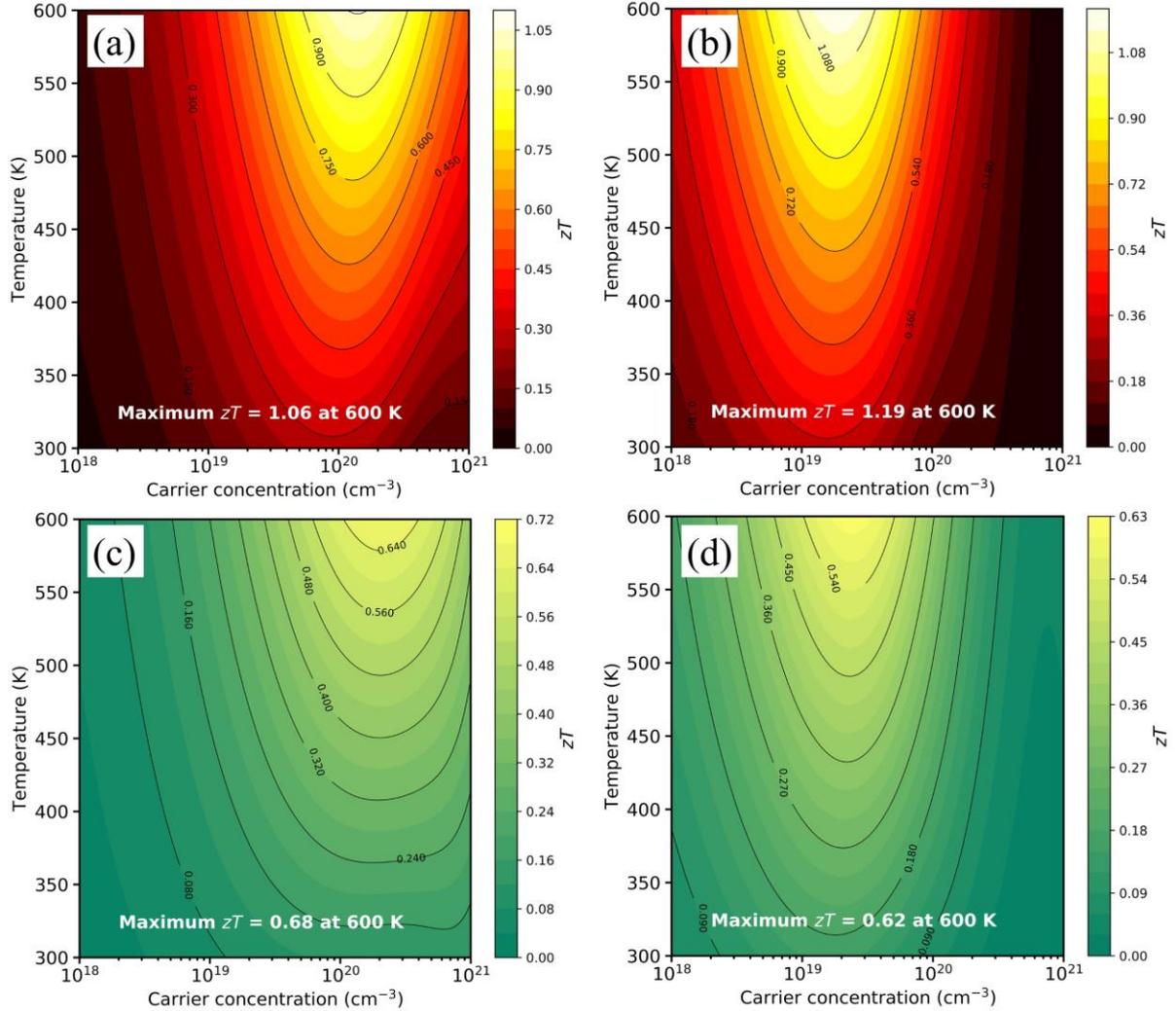

**Figure 9.** Calculated figure of merit of **(a)** p-type, **(b)** n-type $CuBiSeCl_2$ and **(c)** p-type, **(d)** n-type $CuBiSeBr_2$

[63] Based on these results, the study recommends the potential applicability of CuBiSeCl$_2$ system (as both p-type and n-type electrode) for mid temperature thermoelectric energy conversion applications. However, the performance of CuBiSeBr$_2$ needs to be improved further to consider it for prototypical module.

## 4. Conclusions

In summary, this study systematically presents the structural, electronic transport, phonon transport, and thermoelectric properties of the double cation chalcohalides CuBiSeX$_2$ (X = Cl, Br), within the framework of DFT. The experimental synthesizability, as well as the thermal, dynamical, and mechanical stability of the targeted systems have been validated. The structural peculiarities in conjugation with favourable electronic band features lead to enhanced electron transport behaviour. The combined influence of Cu delocalization, flatter potential energy surface, and large atomic displacement parameter facilitates strong Umklapp phonon scattering process. Moreover, the off-centre displacement of Cu and Bi cations from their equilibrium positions promotes bonding anisotropy, giving rise to pronounced lattice anharmonicity. These attributes collectively account for the calculated ultra-low lattice thermal conductivity in both chalcohalides. Quantitatively, $\kappa_L$ of CuBiSeCl$_2$ is lower than CuBiSeBr$_2$, and the underlying origin of this difference is elucidated. Overall, CuBiSeCl$_2$ demonstrates superior thermoelectric performance for both electron and hole doping, whereas the $zT$ of CuBiSeBr$_2$ requires further optimization. In a broad spectrum, the present study incites the exploration and development of double-cation chalcohalide for mid-temperature thermoelectric applications.

**Author contributions**

Manivannan Saminathan: Conceptualization, Software, Formal analysis, Investigation, Writing – original draft; Prakash Govindaraj: Visualization, Methodology, Data curation; Hern Kim: Resources, Project administration, Data curation, Funding acquisition; Kowsalya Murugan: Validation, Writing – review & editing; Kathirvel Venugopal: Supervision, Conceptualization Writing – review & editing.

**Conflict of Interest**

The authors declare no conflict of interest.

**Acknowledgements**

The authors thank High Performance Computing Center, SRM Institute of Science and Technology for providing the computational facility. Also, this work was supported by the National Research Foundation (NRF) grants funded by the Ministry of Education (RS-2020-NR049576), Republic of Korea.

# Supporting Information

# Interplay between Cu Diffusion and Bonding Anisotropy on the thermoelectric performance of double cation chalcohalides CuBiSeX$_2$ (X = Cl, Br)

Manivannan Saminathan[a], Prakash Govindaraj[b], Hern Kim[b], Kowsalya Murugan[a], Kathirvel Venugopal[a, *]

[a]Department of Physics and Nanotechnology, SRM Institute of Science and Technology, Kattankulathur, Tamil Nadu, 603 203, India.

[b]Department of Energy Science and Technology, Environmental Waste Recycle Institute, Myongji University, Yongin, Gyeonggi-do 17058, Republic of Korea.

*Corresponding Author E-mail: kathirvv@srmist.edu.in

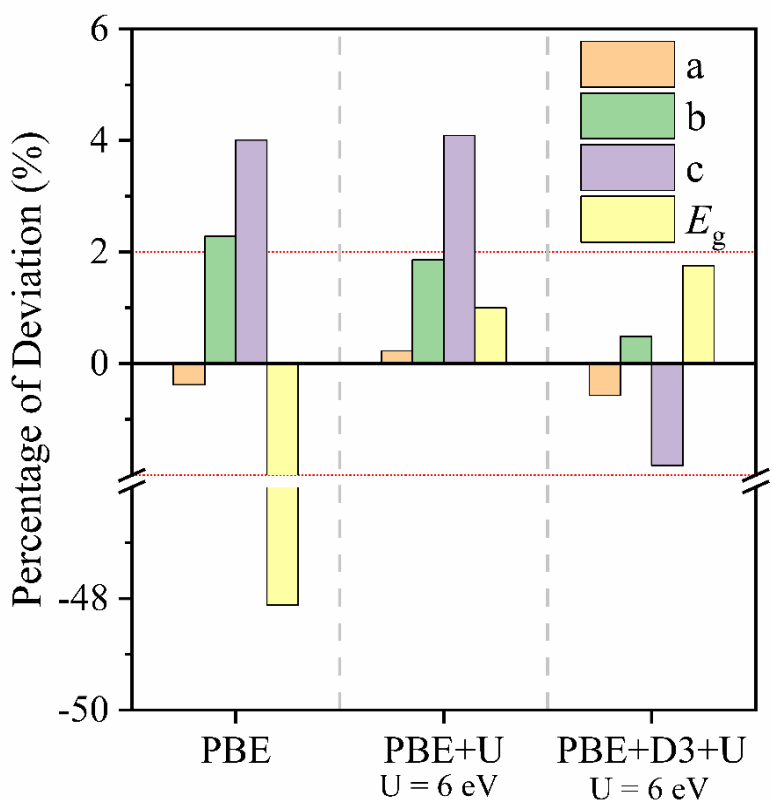

**Figure S1.** Percentage of deviation with experimental report against various exchange correlational function.

**Table S1.** Calculated Static, high-frequency dielectric constant, and polar-phonon frequency.

|  | CuBiSeCl$_2$ | CuBiSeBr$_2$ |
|---|---|---|
| Static dielectric constant matrix | $\begin{bmatrix} 27.3 & 0 & 0 \\ 0 & 26.8 & 0 \\ 0 & 0 & 12.7 \end{bmatrix}$ | $\begin{bmatrix} 37.8 & 0 & 0 \\ 0 & 28.8 & 0 \\ 0 & 0 & 18.1 \end{bmatrix}$ |
| High-frequency dielectric constant matrix | $\begin{bmatrix} 7.3 & 0 & 0 \\ 0 & 17.9 & 0 \\ 0 & 0 & 5.4 \end{bmatrix}$ | $\begin{bmatrix} 9.7 & 0 & 0 \\ 0 & 10.6 & 0 \\ 0 & 0 & 8.3 \end{bmatrix}$ |
| Polar-phonon frequency (in THz) | 3.61 | 2.71 |

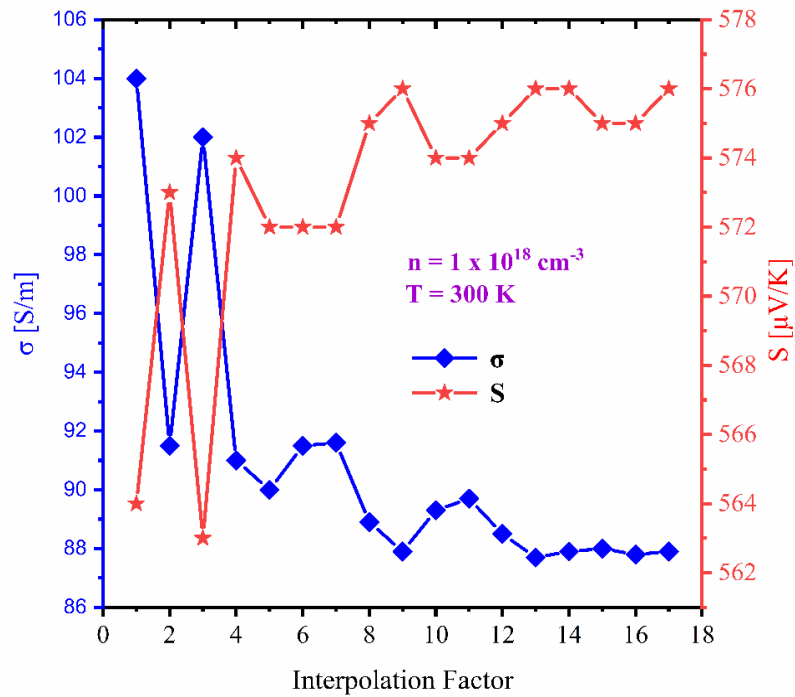

**Figure S2.** AMSET- Interpolation factor convergence test at fixed temperature and carrier concentration.

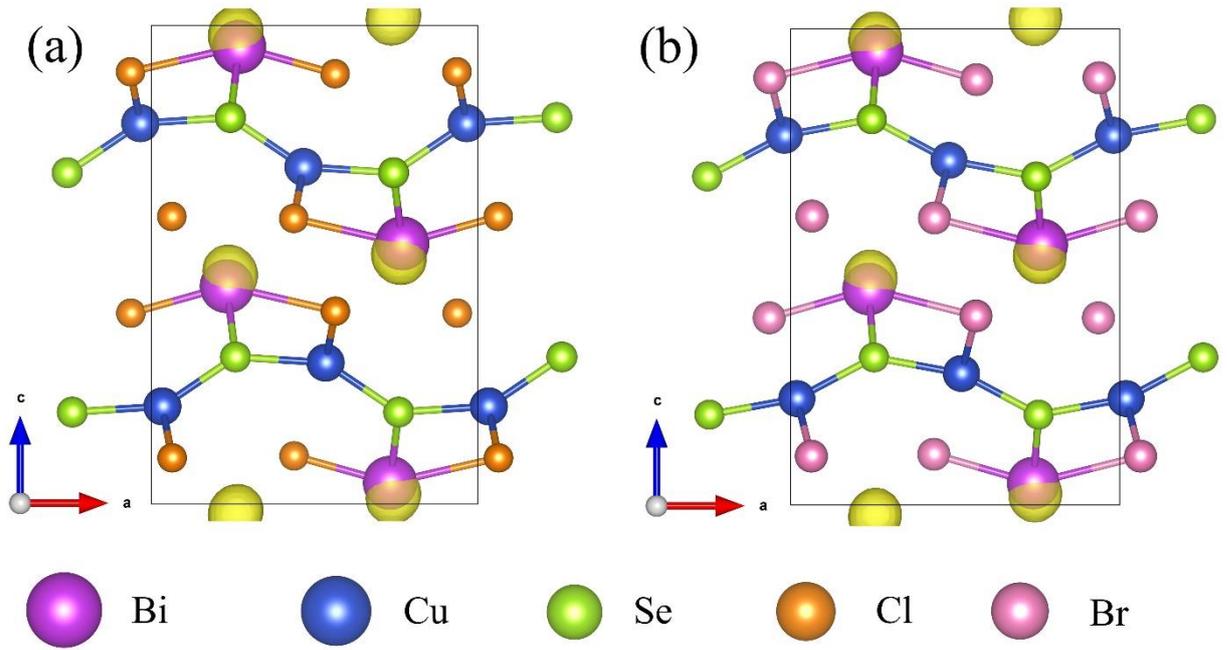

**Figure S3.** The 3D electron localization function of (a) CuBiSeCl2 and CuBiSeBr2 at the isosurface level of 0.92.

**Table S2.** Comparison of the calculated lattice parameter and bandgap of CuBiSeX$_2$ (X= Cl, Br) compounds.

| Material | Lattice Parameter (Å) | | | Band gap (eV) | References |
|---|---|---|---|---|---|
| | *a* | *b* | *c* | | |
| **CuBiSeCl$_2$** | 8.734 | 4.017 | 12.899 | 1.35 | this work |
| | 8.784 | 3.998 | 13.139 | 1.33 | Expt. [1] |
| | 8.73 | 4.05 | 12.99 | 1.30 | Other Calc. [2] |
| **CuBiSeBr$_2$** | 9.063 | 4.094 | 13.204 | 0.85 | this work |

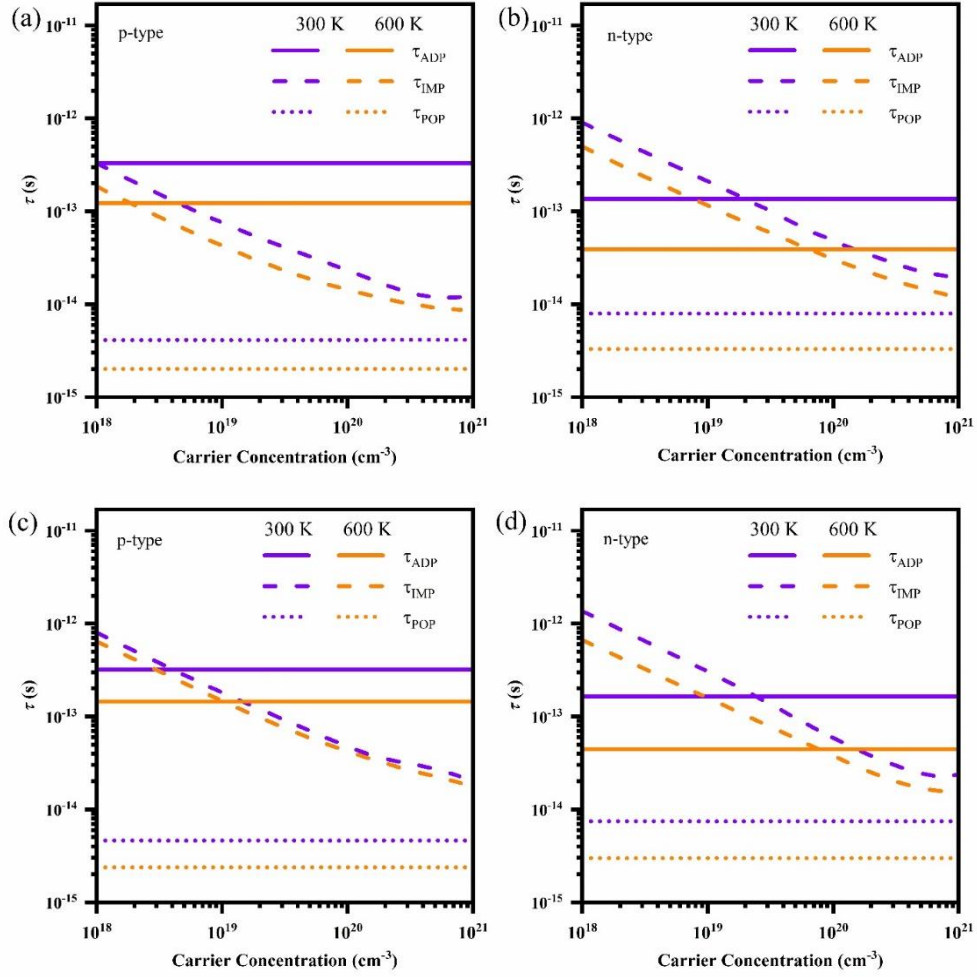

**Figure S4.** The calculated relaxation time contributed by various scattering mechanisms against carrier concentration of (a) p-type CuBiSeCl$_2$, (b) n-type CuBiSeCl$_2$, (c) p-type CuBiSeBr$_2$, and (d) n-type CuBiSeBr$_2$. The dotted and solid lines correspond to 300 and 600 K.

**Table S3** Temperature-dependent total relaxation time ($\tau_{tot}$) for p-type and n-type CuBiSeX$_2$ (X= Cl, Br) for the carrier concentration $1\times10^{19}$ cm$^{-3}$.

|  | Materials | T (K) | p-type | n-type |
|---|---|---|---|---|
| $\tau_{tot}$ (*f*s) | CuBiSeCl$_2$ | 300 | 3.86 | 7.24 |
|  |  | 600 | 1.89 | 2.96 |
|  | CuBiSeBr$_2$ | 300 | 4.43 | 6.98 |
|  |  | 600 | 2.31 | 2.75 |

**Table S4.** Temperature dependent electrical conductivity (σ) for p-type and n-type CuBiSeX$_2$ (X= Cl, Br) for the carrier concentration $1\times10^{19}$ cm$^{-3}$.

|  | Materials | T (K) | p-type | n-type |
|---|---|---|---|---|
| σ (S m$^{-1}$) | CuBiSeCl$_2$ | 300 | 811.35 | 5915.83 |
|  |  | 600 | 380.71 | 2881.56 |
|  | CuBiSeBr$_2$ | 300 | 1527.39 | 7336.75 |
|  |  | 600 | 645.97 | 3773.11 |

**Table S5.** Temperature dependent Seebeck Coefficient (S) for p-type and n-type CuBiSeX$_2$ (X= Cl, Br) at the carrier concentration $1\times10^{19}$ Cm$^{-3}$.

|  | Materials | T (K) | p-type | n-type |
|---|---|---|---|---|
| S (μV K$^{-1}$) | CuBiSeCl$_2$ | 300 | 383 | -228 |
|  |  | 600 | 471 | -309 |
|  | CuBiSeBr$_2$ | 300 | 321 | -218 |
|  |  | 600 | 450 | -291 |

**Table S6.** Comparison of the optimum thermoelectric power factor (PF) and corresponding carrier concentration (*n*) for p-type and n-type CuBiSeX$_2$ (X= Cl, Br).

| Materials | T (K) | p-type | | n-type | |
|---|---|---|---|---|---|
|  |  | PF (mW m$^{-1}$ K$^{-2}$) | n (cm$^{-3}$) | PF (mW m$^{-1}$ K$^{-2}$) | n (cm$^{-3}$) |
| CuBiSeCl$_2$ | 300 | 0.26 | 1.2 x 10$^{20}$ | 0.35 | 2.6 x 10$^{19}$ |
|  | 600 | 0.34 | 4.7 x 10$^{20}$ | 0.44 | 9.0 x 10$^{19}$ |
| CuBiSeBr$_2$ | 300 | 0.50 | 8.6 x 10$^{20}$ | 0.39 | 2.3 x 10$^{19}$ |
|  | 600 | 0.61 | 6.4 x 10$^{20}$ | 0.47 | 5.3 x 10$^{19}$ |

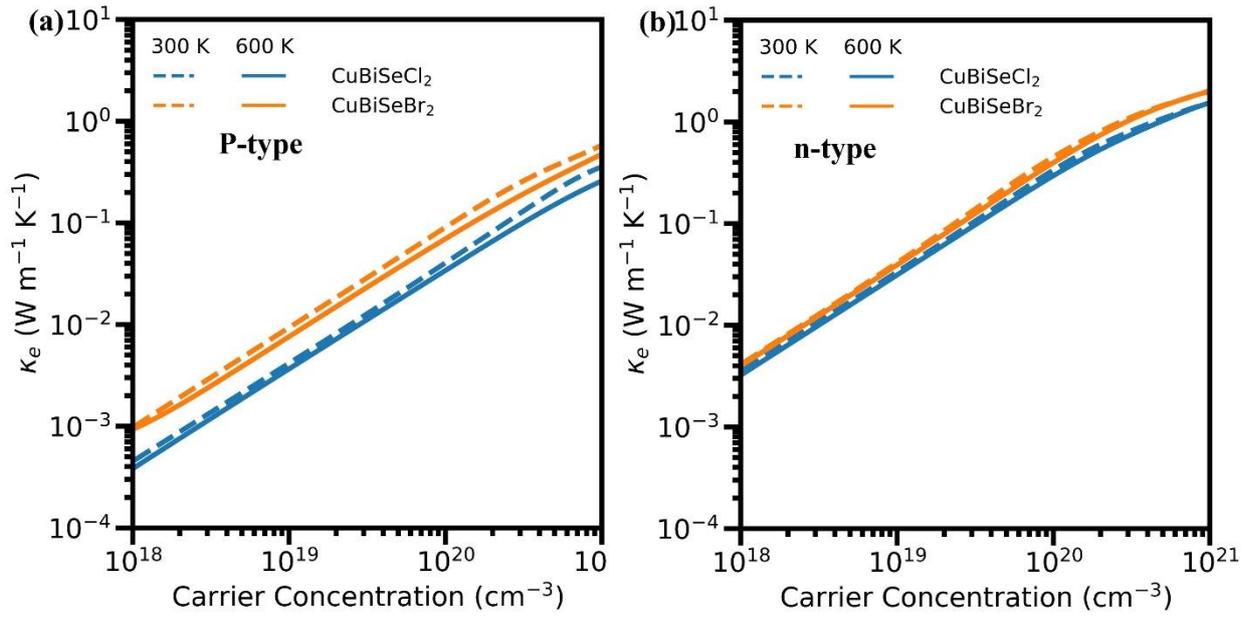

**Figure S5.** The calculated Electronic thermal conductivity against carrier concentration for (a) p-type (b) n-type $CuBiSeX_2$ (X = Cl, Br).

**Table S7.** Temperature dependent electronic thermal conductivity ($\kappa_e$) for p-type and n-type $CuBiSeX_2$ (X= Cl, Br) at the carrier concentration $1 \times 10^{19}$ $Cm^{-3}$.

|  | Materials | T (K) | p-type | n-type |
|---|---|---|---|---|
| $\kappa_e$ (W m$^{-1}$ K$^{-1}$) | $CuBiSeCl_2$ | 300 | 0.004 | 0.033 |
|  |  | 600 | 0.003 | 0.031 |
|  | $CuBiSeBr_2$ | 300 | 0.009 | 0.041 |
|  |  | 600 | 0.007 | 0.039 |

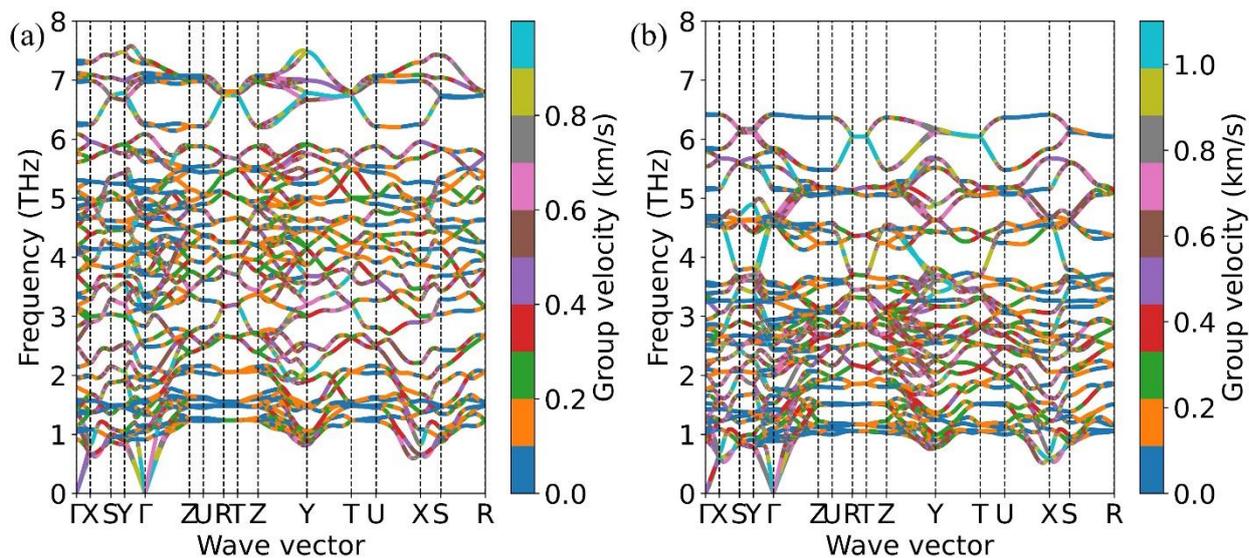

**Figure S6.** The calculated group velocity of **(a)** CuBiSeCl$_2$ and **(b)** CuBiSeBr$_2$.

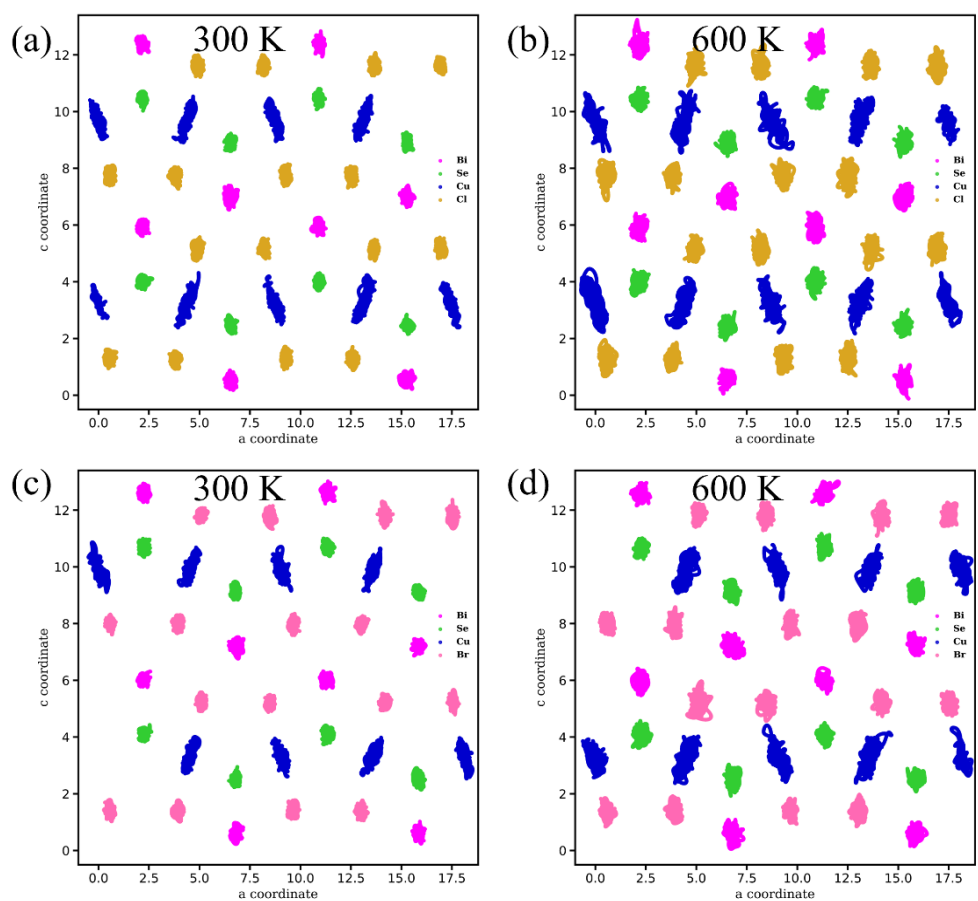

**Figure S7.** The temperature-dependent trajectory of atoms in **(a, b)** CuBiSeCl$_2$ and **(c, d)** CuBiSeBr$_2$ lattice. The blue, magenta, green, orange and pink colors represent copper, bismuth, selenium, chlorine and bromine atom's trajectory.

**Table S8**. Comparison of the optimum thermoelectric Figure of Merit ($zT$) and corresponding carrier concentration ($n$) for p-type and n-type CuBiSeX$_2$ (X= Cl, Br) at 300 and 600 K.

| Materials | T (K) | p-type | | n-type | |
|---|---|---|---|---|---|
| | | $zT$ | n (cm$^{-3}$) | $zT$ | n (cm$^{-3}$) |
| CuBiSeCl$_2$ | 300 | 0.28 | 8.7 x 10$^{19}$ | 0.34 | 1.6 x 10$^{19}$ |
| | 600 | 1.05 | 1.4 x 10$^{20}$ | 1.18 | 1.9 x 10$^{19}$ |
| CuBiSeBr$_2$ | 300 | 0.13 | 7.1 x 10$^{20}$ | 0.15 | 1.7 x 10$^{19}$ |
| | 600 | 0.68 | 2.1 x 10$^{20}$ | 0.61 | 2.4 x 10$^{19}$ |